\documentstyle[12pt]{article}
\begin{document}

\centerline{\bf Pole versus Breit-Wigner resonance} 
\centerline{\bf description of the orbitally excited baryons} 
\vskip .3cm
\centerline{Ron Workman} 
\centerline{Department of Physics, Virginia Tech, Blacksburg, VA 24061 }

\begin{abstract}

We consider the masses used in recent studies involving the nonstrange
sector of the $l$ = 1 baryons. The use of $T$-matrix and $K$-matrix poles 
versus the conventional Breit-Wigner masses
is discussed within the context of a large-$N_c$ fitting scheme.

\end{abstract}

\vskip .5cm

While the mass and charge of a particle are typically the easiest quantities to 
determine experimentally, the proper way to define and extract the mass of an
unstable state continues to be controversial. 
This issue has been extensively debated in
studies of the $Z^0$ mass\cite{Zpole} and has been discussed, in the context
of Baryon resonances, by H\"ohler\cite{Hoehler}. While resonance-like 
behavior is, in principle, possible without a pole in the 
$S$-matrix\cite{Fonda}, the pole position has many features one would associate
with the physical mass. These include\cite{Zpole,properties} independence from 
the production process, factorizability of the residue, and gauge 
independence. Other possibilities include the bare, $K$-matrix, and 
Breit-Wigner (BW) masses.

The Particle Data Group (PDG) has until recently listed only the
BW masses and widths in its Baryon Summary Tables\cite{PDG}, though 
pole positions have been added in the most recent edition. Most fits 
involving either quark model or large-$N_c$ formalisms have been carried out
using these BW values. However, as emphasized by H\"ohler\cite{Hoehler}, 
the BW values quoted by the PDG are inherently model-dependent. This has led
us to ask two important questions. (1) Are the above fits influenced by 
differences between the BW and pole masses? (2) Which definitions of the mass
actually correspond to the quantities being calculated?

The first question can be answered most easily. For this purpose, we have
examined several fits to resonance properties utilizing the large-$N_c$ 
formalism. In a series of papers, the orbitally excited
SU(6) {\bf 70}-plet baryons have been analyzed in terms of their BW
masses\cite{ln1}, and strong\cite{ln2} and electromagnetic\cite{ln3}
decays, within the framework of large-$N_c$ QCD. 
While such fits necessarily involve a large number of free parameters,
a comparison of the parameters determined in these independent fits
reveals a remarkably consistency. This is particularly evident if one
compares the mixing angles associated with, in $\pi N$ notation, the
$S_{11}$ and $D_{13}$ resonances. The two $S_{11}$ mass eigenstates, 
N(1535) and N(1650), and two $D_{13}$ mass eigenstates, N(1520) and
N(1700), are mixtures of states, $N_{ij}$, with total 
quark spin $i/2$ and total angular momentum $j/2$, as parameterized
by mixing angles:
\begin{equation}
\left[ \begin{array}{c}
N(1535) \\
N(1650) \end{array} \right]
=
\left[ \begin{array}{cc}
\cos \theta_{N1} & \sin \theta_{N1} \\
-\sin \theta_{N1} & \cos \theta_{N1} \end{array} \right]
\left[ \begin{array}{c}
 N_{11} \\
 N_{31} \end{array} \right]
\end{equation}
and
\begin{equation}
\left[ \begin{array}{c}
N(1520) \\
N(1700) \end{array} \right]
=
\left[ \begin{array}{cc}
\cos \theta_{N3} & \sin \theta_{N3} \\
-\sin \theta_{N3} & \cos \theta_{N3} \end{array} \right]
\left[ \begin{array}{c}
 N_{13} \\
 N_{33} \end{array} \right] .
\end{equation}
The mixing angles, $\theta_{N1}$ and $\theta_{N3}$, have been determined
independently in Refs.\cite{ln1,ln2,ln3}. Results for the
angles are identical, within the quoted uncertainties, in these
fits to the masses and
decay widths (both strong and electromagnetic). This self-consistency
adds considerable weight to the large-$N_c$ fitting scheme. 

We have repeated the mass fit of Ref.\cite{ln1} using instead a 
set of pole masses\cite{com1,pin,Green}, where the mass was taken to
to be the real part of the pole position. For these resonances, the
difference in definition is quite important. This is apparent
if one notes that the heaviest and lightest of the orbitally excited
SU(6) {\bf 70}-plet baryons are separated by only about 200 MeV, whereas
the difference between BW and pole masses can be 50 MeV or more.

While any comparison necessarily depends upon the
number of operator coefficients varied in the fit, an interesting result
follows if one fits the seven masses with six parameters and predicts the
two mixing angles. These parameters scale the O($N_c$) and O(1)
contributions and the largest terms of O($N_c^{-1}$). 
A detailed description of this method and a complete set of relations 
between the parameters and masses are given in Ref.\cite{ln1}. 
In the present short note, we retain this notation\cite{ln1}
in order to aid comparison.  The six-parameter fit of Ref.\cite{ln1} 
was able to successfully reproduce the BW masses and 
mixing angles in agreement with the results of Refs.\cite{ln2,ln3}. 
A nine-parameter fit, including the mixing angles as data, did not give
qualitatively different results\cite{ln1}. (The values
given in Ref.\cite{ln1} were first verified before considering the 
effect of pole masses\cite{com}.) Our fits, 
using both BW and pole masses, and the resulting parameters and
mixing angles are given in Tables I and II.

While the mixing angles resulting from the pole fit are quite different
from those found in Refs.\cite{ln1,ln2,ln3}, the other parameters 
display a number of similarities. The relative
signs have not changed and the coefficient $c_2$ remains small
and consistent with zero. Apart from $c_2$, the coefficients found 
in the pole-mass fit are of the same magnitude.
The other terms of O($N_c^{-1}$)
listed and considered in Ref.\cite{ln1} appear to be unimportant
in both the BW and pole-mass fits. It is also interesting to see that,
using either set of masses, a significant part of the overall chi-squared 
is due to the $N(1700)$. This state has a very weak coupling to the 
$\pi N$ channel, and has not been detected in all analyses of elastic
$\pi N$ scattering data. As a result, its mass and pole position are
not well determined. 

In an expanded version of Ref.\cite{ln1}, the relative sizes of the fitted
coefficients have been used to suggest that the underlying dynamics is due to 
effective pseudoscalar-meson exchanges among the quarks\cite{ln4}. 
As the pole mass fit chooses a different set of dominant coefficients, 
we see that differences in definition {\it are} important. Thus we are 
forced to consider the second (much harder) question: Which ``mass'' is
most appropriate?

We first assume that the pole positions are eigenvalues of an operator
$M- {i\over 2} \Gamma$ in the sense that
\begin{equation}
\left( M -  {i\over 2} \Gamma \right) | A > = m_A | A >
\end{equation}
where $m_A$ is complex. The connection with most model approaches is a
neglect of $\Gamma$, resulting in real mass values. For the states under
consideration, terms of order $N_c$, 1, and $N_c^{-1}$ have been included
in fits to the masses. As the widths are expected to enter at 
O($N_c^{-2}$)\cite{Wirzba}, the neglect of $\Gamma$ appears to be 
completely consistent.

In order to more closely examine the mixed states N(1535) and N(1650), 
or N(1520) and N(1700), we follow an argument given by 
Aitchison\cite{Aitchison} for overlapping resonances. In this case,
$M- {i\over 2}\Gamma$ is an effective Hamiltonian matrix. In terms of
states, denoted by Greek indices, which diagonalize $M$ 
(but not the full Hamiltonian), the 
$T$-matrix has the form\cite{Aitchison}
\begin{equation}
T_{ij} = f_{i \alpha } S_{\alpha \beta }' f_{\beta j}
\end{equation}
where  
\begin{equation}
S_{\alpha \beta }^{' -1} = S_{\alpha \beta }^{-1} - \Sigma_{\alpha \beta }
\end{equation}
and
\begin{equation}
S_{\alpha \beta }^{-1} = \left( m_{\alpha } - m \right) 
\delta_{\alpha \beta },
\;\;\; \Sigma_{\alpha \beta } = {i\over 2} \sum_i 2\pi \rho_i f_{\alpha i}
f_{i \beta } .
\end{equation}
In the above, $m$ is the energy, $f_{\alpha i}$ is the coupling between the
resonant state $\alpha$ and the continuum state $i$, and $\rho_i$ is the
phase space factor for channel $i$. The $T$-matrix can then be written, in
terms of the $K$-matrix, as $T=K(1-i \pi \rho K )^{-1}$, wherein the
$K$-matrix has the form
\begin{equation}
K_{ij} = f_{i \alpha } S_{\alpha \beta } f_{\beta j} .
\end{equation}
The neglect of $\Gamma$ results in the approximation $T_{ij} \simeq K_{ij}$.
Therefore, as one might expect, $K$- and $T$-matrix poles are equivalent in
the absence of a width. 

At this point a few comments are necessary. First, it is known that the
$K$- and $T$-matrix masses are separated by amounts similar to the 
difference between the BW and $T$-matrix pole masses\cite{Green,Long}.
Thus, one or both of the $K$- and $T$-matrix masses must shift significantly
in the presence of a width. From the fits in Tables I and II, 
we see that the masses can be reproduced, to the few MeV level, without
O($N_c^{-2}$) terms. As a result, we expect the $K$-matrix mass to 
remain relatively stable. However, in Eq.~(3)
we see that a width alters both the effective Hamiltonian and the
basis states. As this width is not small, being typically 150 MeV,
a moderate shift in the real part of $m_A$ should be expected. 
In matching phenomenological masses from data fits to the formalism of
Refs.\cite{ln1,ln4}, we require a quantity which remains 
stable as the width is turned on.  As a result we suggest that the (real) 
$K$-matrix poles are most closely
associated with the large-$N_c$ formalism of Refs.\cite{ln1,ln2,ln3,ln4}.
While $K$-matrix pole positions are not tabulated by the PDG, a recent
study\cite{Green} finds that, at least for the N(1535), the BW and $K$-matrix 
masses are in reasonable agreement.  

In summary, after comparing the various definitions used to extract masses from
experimental data, at least for the considered set of resonances, we
find the $K$-matrix definition to be most appropriate when comparing with 
large-$N_c$ results. One might object that large-$N_c$ QCD should be giving
the more physical $T$-matrix result. This is not a problem, as the above 
argument implies that a comparison of phenomenological $K$-matrix masses is
essentially equivalent to a comparison of $T$-matrix masses to O($N_c^{-1}$). 
One final point should be emphasized.
In this study, we have completely ignored the effects
of non-resonant background contributions. This would not have been a
problem had the $T$-matrix pole been favored. However, the $K$-matrix pole
is influenced by background contributions, and thus a degree of 
model-dependence appears unavoidable\cite{com_Kmatrix}. As a final point,
we mention that some recent studies\cite{data} have found photo-decay
amplitudes for the N(1535) and N(1520) which strongly contradict the PDG
values fitted in Ref.\cite{ln3}. In the author's opinion, it would be
extremely useful to determine whether these results preserve the consistent
picture found in Refs.\cite{ln1,ln2,ln3,ln4}.

We thank Carl Carlson, Chris Carone, and Tetsuro Mizutani for many helpful 
discussions.  A useful communication from H.B.~O'Connell is also acknowledged.
This work was supported in part by a U.S. Department of Energy Grant No.
DE-FG02-97ER41038.

\newpage
\begin{table}
\caption{ Six parameter fit to BW masses. The predicted mixing angles 
are: $\theta_{N1}$=0.53 radians and $\theta_{N3}$=3.06 radians. 
(Values from Ref.\cite{ln2} are:
$\theta_{N1}$=0.61$\pm$0.09 radians and 
$\theta_{N3}$=3.04$\pm$0.15 radians.) 
The $\chi^2$/d.o.f.=0.23. Mixing angles were not included as data.
}
\label{table1}
\begin{center}
\begin{tabular}{|l|l|l|l|}\hline
 & Fit (MeV) & Exp. (MeV) & Parameters (MeV)\\ \hline \hline
$\Delta (1700)$ & 1712 & 1720 $\pm$ 50 & $c_1$: 466 $\pm$ 14      \\
$\Delta (1620)$ & 1643 & 1645 $\pm$ 30 & $c_2$: -29.5 $\pm$ 39     \\
$N(1675)$       & 1678 & 1678 $\pm$ 8  & $c_3$: 303 $\pm$ 141     \\
$N(1700)$       & 1712 & 1700 $\pm$ 50 & $c_4$: 69 $\pm$ 99     \\
$N(1650)$       & 1660 & 1660 $\pm$ 20 & $c_5$: 63 $\pm$ 46     \\
$N(1520)$       & 1523 & 1523 $\pm$ 8 & $c_6$: 424 $\pm$ 86     \\
$N(1535)$       & 1539 & 1538 $\pm$ 18 &      \\ \hline \hline
\end{tabular}
\end{center}
\end{table}

\begin{table}
\caption{ Six parameter fit to pole masses. The predicted mixing angles
are: $\theta_{N1}$=2.63 radians and  $\theta_{N3}$=0.35 radians.
The $\chi^2$/d.o.f.=0.005. Mixing angles were not included as data.
}
\label{table2}
\begin{center}
\begin{tabular}{|l|l|l|l|}\hline
 & Fit (MeV) & Exp. (MeV) & Parameters (MeV)\\ \hline \hline
$\Delta (1700)$ & 1655 & 1655 $\pm$ 10 & $c_1$: 497 $\pm$ 13      \\
$\Delta (1620)$ & 1585 & 1585 $\pm$ 15 & $c_2$: -1.7 $\pm$ 18     \\
$N(1675)$       & 1660 & 1660 $\pm$ 10 & $c_3$: 196 $\pm$ 87     \\
$N(1700)$       & 1647 & 1650 $\pm$ 50 & $c_4$: 186 $\pm$ 26     \\
$N(1650)$       & 1670 & 1670 $\pm$ 20 & $c_5$: 104 $\pm$ 21     \\
$N(1520)$       & 1510 & 1510 $\pm$ 5  & $c_6$: 212 $\pm$ 70     \\
$N(1535)$       & 1510 & 1510 $\pm$ 10 &      \\ \hline \hline
\end{tabular}
\end{center}
\end{table}


\begin{thebibliography}{99}
\bibitem{com1}
The pole masses fitted in Table II were estimated from the results
of Refs.\cite{PDG,pin,Green}. The N(1700) state was not detected
in the analysis of Ref.\cite{pin}.
\bibitem{com}
The Tables of coefficients and chi-squared values were reproduced 
after many invaluable communications with 
the authors of Ref.\cite{ln1}.
\end{thebibliography}
\end{document}